\documentclass[%
 aip,
 amsmath,amssymb,
 reprint,%
floatfix,%
]{revtex4-1}

\usepackage{graphicx}
\usepackage{dcolumn}
\usepackage{bm}

\usepackage[utf8]{inputenc}
\usepackage[T1]{fontenc}
\usepackage{mathptmx}
\usepackage[compact]{titlesec}
\usepackage{xcolor}

\begin{document}

\preprint{AIP/123-QED}

\title[Permittivity and permeability of epoxy-magnetite powder composites at microwave frequencies]{Permittivity and permeability of epoxy-magnetite powder composites at microwave frequencies}

\author{T. Ghigna}
\altaffiliation[Author to whom correspondence should be addressed:]{ tommaso.ghigna@physics.ox.ac.uk}
\affiliation{\mbox{Sub-Department of Astrophysics, University of Oxford, Keble Rd, Oxford OX1 3RH, United Kingdom}}
\affiliation{\mbox{Kavli IPMU (WPI), UTIAS, The University of Tokyo, Kashiwa, Chiba 277-8583, Japan}}

\author{M. Zannoni}
\affiliation{\mbox{Dipartimento di Fisica, Universit\'a di Milano-Bicocca, P.za della Scienza 3, 20126, Milano, Italy}}
\affiliation{\mbox{INFN, Sez. di Milano-Bicocca, P.za della Scienza 3, 20126, Milano, Italy}}

\author{M.E. Jones}
\affiliation{\mbox{Sub-Department of Astrophysics, University of Oxford, Keble Rd, Oxford OX1 3RH, United Kingdom}}

\author{A. Simonetto}
\affiliation{\mbox{CNR - Istituto per la Scienza e la Tecnologia dei Plasmi (ISTP-MI), via R. Cozzi 53, 20125, Milano, Italy}}

\date{\today}

\begin{abstract}
Radio, millimetre and sub-millimetre astronomy experiments as well as remote sensing applications often require castable absorbers with well known electromagnetic properties to design and realize calibration targets. In this context, we fabricated and characterized two samples using different ratios of two easily commercially available materials: epoxy (Stycast 2850FT) and magnetite ($\mathrm{Fe_{3}O_{4}}$) powder. We performed transmission and reflection measurements from 7 GHz up to 170 GHz with a VNA equipped with a series of standard horn antennas. Using an empirical model we analysed the data to extract complex permittivity and permeability from transmission data; then we used reflection data to validate the results.
In this paper we present the sample fabrication procedure, analysis method, parameter extraction pipeline, and results for two samples with different epoxy-powder mass ratios.
\end{abstract}

\maketitle

\section{\label{sec:intro}INTRODUCTION}
Castable electromagnetic wave absorbers are usually composite materials made with a polymer encapsulant matrix and a dielectric or magnetic filler. There are examples of this type of material commercially available. The vendors provide electromagnetic properties (dielectric permittivity, magnetic permeability or loss tangents) up to $\sim 18-20$ $\mathrm{GHz}$\cite{eccosorb}, although, for many applications, values at higher frequencies are required to help instrument designs\cite{CalTerget1, CalTarget2, CalTarget3, CalTarget4, CalTerget5}. Examples of instruments with custom designed loads include the Atacama Large Millimeter Array\cite{ALMA, AlmaTarget1, AlmaTarget2} (ALMA), the Planck Low Frequency Instrument\cite{lfiPlanck, lfiLoad} (LFI) and the proposed Primordial Inflation Explorer\cite{pixie} (PIXIE) as examples concerning astrophysical experiments, or the European Space Agency (ESA) MetOp\cite{MetOp, MetOpTarget1, MetOpTarget2} satellite project for what concerns remote sensing applications.

In this paper, we explore the possibility of fabricating a castable absorber using cheap commercially available materials: Stycast 2850 FT\cite{stycast} as the dielectric encapsulant, and magnetite powder\cite{magnetite} (chemical composition $\mathrm{Fe_{3}O_{4}}$) as the magnetic filler, and we measure the properties up to $170$ $\mathrm{GHz}$ with a vector network analyser (VNA).

\begin{figure}[htbp]
    \centering
    \includegraphics[width=0.45\textwidth]{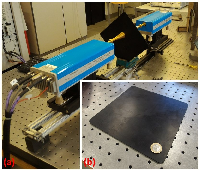}
    \caption{a) Measurement set-up. Two horns antenna are facing each other with the thin flat sample on the aperture of one of the antennas. Transmission and reflection data are measured using a VNA. b) One of the measured samples with a 1 euro coin for size reference.}
    \label{fig:setup}
\end{figure}

We made two samples with different encapsulant-filler mass ratio, in order to show the possibility of tuning the properties depending on a specific experiment needs. The maximum magnetite particle size is 45 $\mu$m and the typical size is 200 nm. We do not expect the particle size to make a significant difference as it is very small compared to the wavelength. We measured transmission and reflection from the samples below 170 GHz, corresponding to wavelengths $\gtrsim 1.7$ mm.
We measured transmittance and reflectance through a thin ($\sim 2$ mm) slab of material, and we used an empirical model to extract information about the electromagnetic properties of our samples through transmission data\cite{ZivMu1, ZivMu2, ZivMu3}. The measurement set-up and one of the measured samples can be seen in Fig. \ref{fig:setup}. To validate our results we use the extracted parameters to calculate, by means of an empirical model, reflection data and we compare them with measurements.

In Sec. \ref{sec:abs} we present the model and the method for the electromagnetic parameters retrieval and in Sec. \ref{sec:res} we present and discuss the results obtained for the samples under study.

\begin{table}[htbp]
    \centering
    \begin{tabular}{c c c c c c c}
    \hline
    Sample & \mbox{ } & Mass-ratio & \mbox{ } & $\mathrm{\Delta x}$ & \mbox{ } & $\mathrm{\epsilon_{dc}}$ \\
    \hline
    \textit{Mag27} & \mbox{ } & 27\% & \mbox{ } & 2.38 $\mathrm{mm}$ & \mbox{ } & 7.57 \\
    \textit{Mag60} & \mbox{ } & 60\% & \mbox{ } & 2.31 $\mathrm{mm}$ & \mbox{ } & 17.79 \\
    \hline
    \end{tabular}
     \caption{Summary of the samples analysed in this study. Powder-to-epoxy mass-ratio, thickness and permittivity measured with the capacitor are shown for each sample.}
    \label{tab:samples}
\end{table}

\section{\label{sec:abs}CHARACTERIZATION OF AN ABSORBER}
In general, we describe the behaviour of matter in presence of an electromagnetic field through the complex relative dielectric permittivity and magnetic permeability. Often we can assume these quantities to be constant, especially if we limit our analysis to a narrow frequency band. But this approximation is not always valid when we extend the analysis to a broader frequency range. In this case we have to take into account a direct dependence from the frequency:
\begin{equation}\label{eq:eps}
    \mathrm{\epsilon(\nu) = \epsilon_r(\nu)+i\epsilon_i(\nu)},
\end{equation} 
\begin{equation}\label{eq:mu}
\mu(\nu) = \mu_r(\nu)+i\mu_i(\nu).
\end{equation} 

\subsection{\label{sec:abs2}PARAMETERS EXTRACTION}
Using Eq. \ref{eq:eps} and \ref{eq:mu}, we can write the complex frequency-dependent refractive index: 
\begin{equation}\label{eq:1}
    \mathrm{\hat{n}(\nu) = \sqrt{\epsilon(\nu)\mu(\nu)} = n(\nu)+ik(\nu).}
\end{equation}
From Eq. \ref{eq:1} it is easy to find the key parameters to compute reflection and transmission through the medium\cite{Yang2011, Sung-Soo}, the reflection parameter (Eq. \ref{eq:2}) and $\delta$ which combine the phase shift and damping factor (Eq. \ref{eq:3}).
\begin{equation}\label{eq:2}
    \mathrm{R = \Big|\frac{\hat{n}(\nu)-1}{\hat{n}(\nu)+1}\Big|^2}
\end{equation}
\begin{equation}\label{eq:3}
    \mathrm{\delta \propto \exp\Big(\frac{i2\pi\hat{n}(\nu)\nu\Delta x}{c}\Big)}
\end{equation}
Using a Debye relaxation model\cite{Sihvola} for the relative permittivity of the medium 
\begin{equation}\label{eq:4}
    \mathrm{\epsilon(\nu) = \epsilon_{\infty} + \frac{\epsilon_{dc}-\epsilon_{\infty}}{1-i\frac{\nu}{\nu_{\epsilon}}}},
\end{equation}
and a Lorenzian resonant model for the relative permeability \begin{equation}\label{eq:5}
     \mathrm{\mu(\nu) = 1 + \frac{\mu_{dc}-1}{\Big(1-i\frac{\nu}{\nu_{\mu}}\Big)^2}}
\end{equation}
we can describe the transmission and reflection from a slab of material with known thickness $\mathrm{\Delta x}$ with five free parameters: $\mathrm{\epsilon_{\infty}}$ (relative permittivity at high frequency), $\mathrm{\epsilon_{dc}}$ (static relative permittivity), $\mathrm{\nu_{\epsilon}}$ (relaxation frequency), $\mathrm{\mu_{dc}}$ (static relative permeability) and $\mathrm{\nu_{\mu}}$ (resonant frequency).

\subsection{\label{sec:fab}SAMPLE FABRICATION}
To perform the analysis we fabricate thin ($\mathrm{\Delta x}\sim 2$ $\mathrm{mm}$) square samples ($200\times200$ $\mathrm{mm}$). To fabricate the samples we built a mould consisting of a flat 10-mm aluminium base plate covered with a 1-mm PTFE sheet to facilitate the release of the finished sample. This PTFE sheet is kept in place by a second 10-mm aluminum plate with a square aperture ($200\times200$ $\mathrm{mm}$) which can be screwed to the base plate to form the moulding structure. To obtain the required thickness we made a PTFE lid, to be inserted into the mould aperture, of proper dimensions to obtain the $\sim 2$ $\mathrm{mm}$ thick sample. Once the epoxy-powder mixture is poured into the mould, and the lid is inserted, the exceeding material can flow out through 4 holes at the corners of the lid. We used a release agent to prevent the sample from sticking to the aluminum walls of the mould.

In this paper we present the results of two samples made using Stycast 2850FT (ideal for cryogenic applications), catalyst 24 LV (7\% mass ratio) and $\mathrm{Fe_3O_4}$-powder. Both samples were prepared by mixing the components (epoxy, catalyst and magnetite powder), outgassing the mixture to ensure a uniform sample, and curing the sample in a oven at $65^{\mathrm{o}}\mathrm{C}$.

The two samples differ in mass-ratio between the magnetic powder and the epoxy encapsulant: 27\% and 60\%. From now on we will refer to the first material as {\it Mag27} and the second as {\it Mag60}.

\subsection{\label{sec:abs1}MEASUREMENT SET-UP}
\begin{figure}[htbp]
    \centering
    (a)\includegraphics[width=0.45\textwidth]{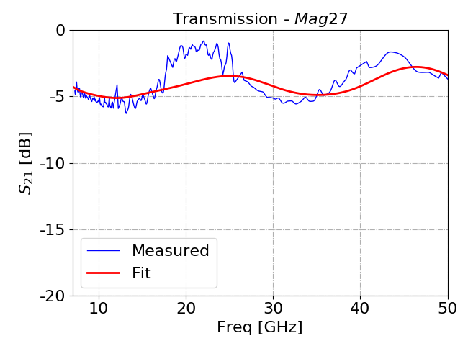}
    (b)\includegraphics[width=0.45\textwidth]{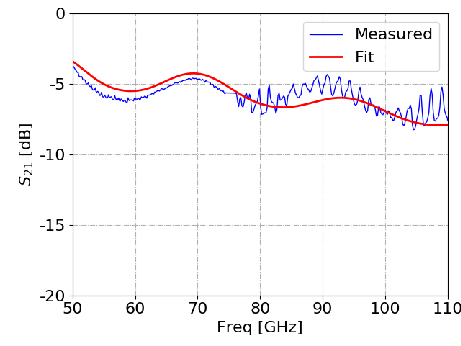}
    (c)\includegraphics[width=0.45\textwidth]{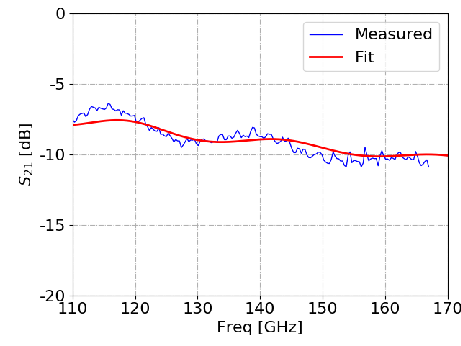}
    \caption{Transmission data and fit for {\it Mag27}. Data and the result of the analysis are split in 3 sub-plots for clarity: a) X, Ku, K, Ka and Q bands, b) V and W bands, c) D band.}
    \label{fig:mag27_trans3}
\end{figure}
\begin{figure}[htbp]
    \centering
    (a)\includegraphics[width=0.45\textwidth]{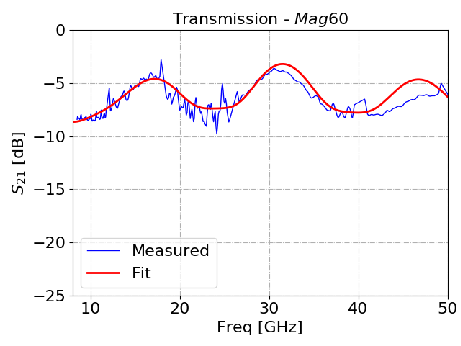}
    (b)\includegraphics[width=0.45\textwidth]{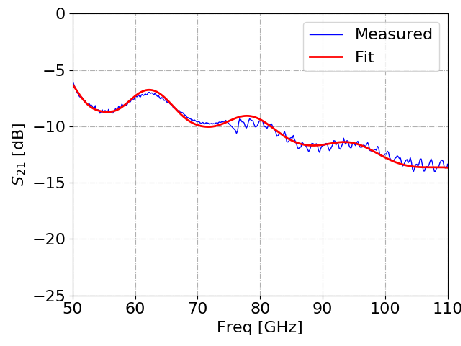}
    (c)\includegraphics[width=0.45\textwidth]{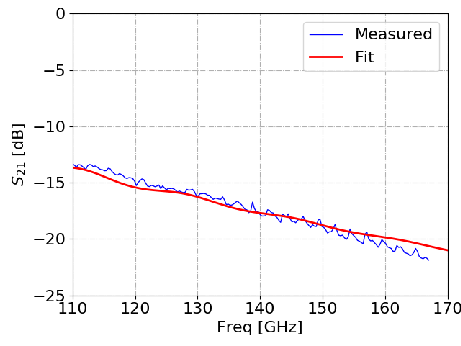}
    \caption{Transmission data and fit for {\it Mag60}. Data and the result of the analysis are split in 3 sub-plots for clarity: a) X, Ku, K, Ka and Q bands, b) V and W bands, c) D band.}
    \label{fig:mag60_trans3}
\end{figure}
\begin{figure}[htbp]
    \centering
    (a)\includegraphics[width=0.45\textwidth]{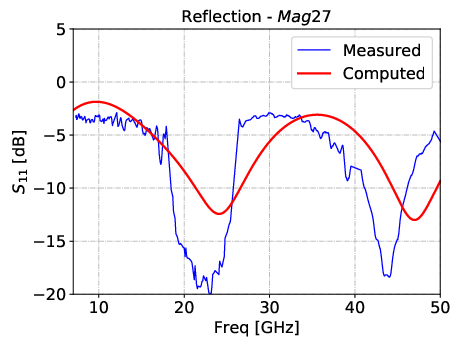}
    (b)\includegraphics[width=0.45\textwidth]{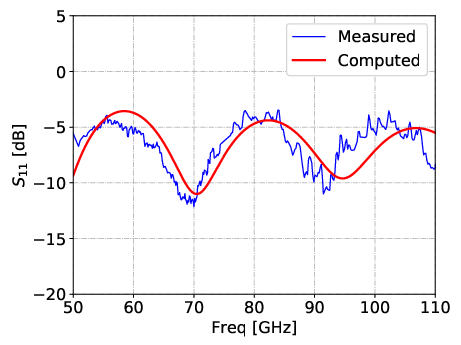}
    (c)\includegraphics[width=0.45\textwidth]{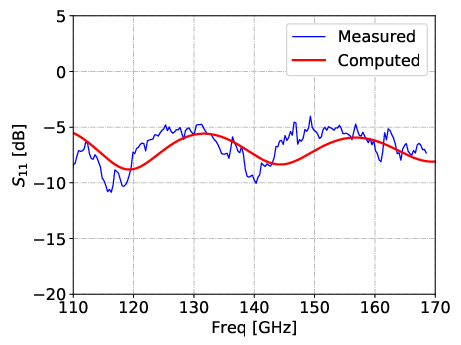}
    \caption{Reflection data and simulated data based on the extracted permittivity and permeability for {\it Mag27}. Data and the result of the analysis are split in 3 sub-plots for clarity: a) X, Ku, K, Ka and Q bands, b) V and W bands, c) D band.}
    \label{fig:mag27_ref3}
\end{figure}
\begin{figure}[htbp]
    \centering
    (a)\includegraphics[width=0.45\textwidth]{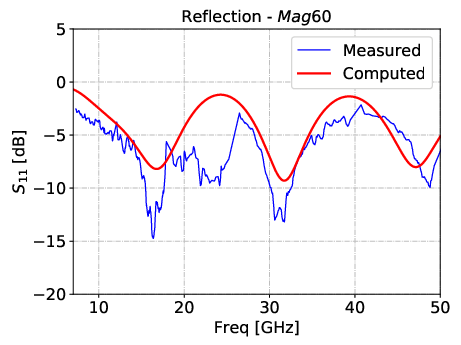}
    (b)\includegraphics[width=0.45\textwidth]{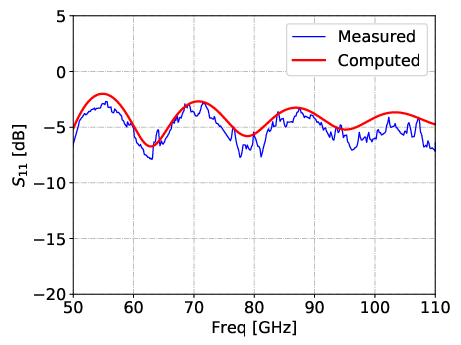}
    (c)\includegraphics[width=0.45\textwidth]{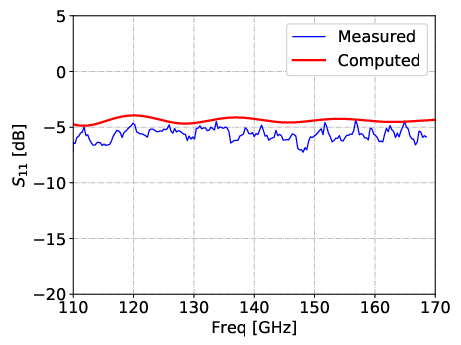}
    \caption{Reflection data and simulated data based on the extracted permittivity and permeability for {\it Mag60}. Data and the result of the analysis are split in 3 sub-plots for clarity: a) X, Ku, K, Ka and Q bands, b) V and W bands, c) D band.}
    \label{fig:mag60_ref3}
\end{figure}
\begin{figure}[htbp]
    \centering
    (a)\includegraphics[width=0.45\textwidth]{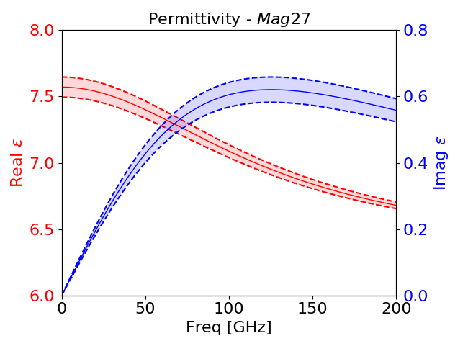}
    (b)\includegraphics[width=0.45\textwidth]{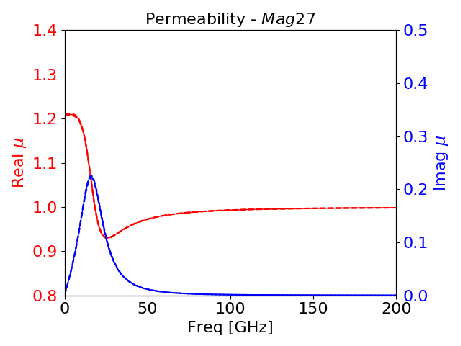}
    \caption{Dielectric permittivity and magnetic permeability for \textit{Mag27}, the real parts (\textit{Real $\epsilon$} and \textit{Real $\mu$)} are plotted in \textit{red}, while the imaginary part \textit{Imag $\epsilon$} and \textit{Imag $\mu$}) are plotted in \textit{blue}. The \textit{solid line} represents the best fit, while the \textit{shadowed area} between the \textit{dashed lines} shows the $1-\sigma$ uncertainty after the fitting routine.}
    \label{fig:mag27}
\end{figure}
\begin{figure}[htbp]
    \centering
    (a)\includegraphics[width=0.45\textwidth]{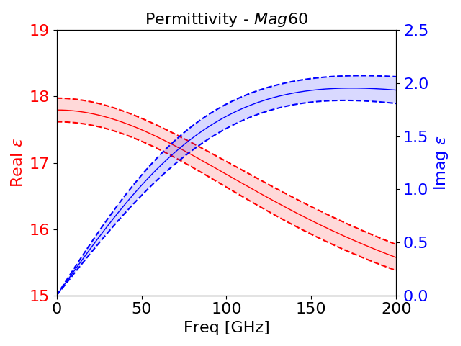}
    (b)\includegraphics[width=0.45\textwidth]{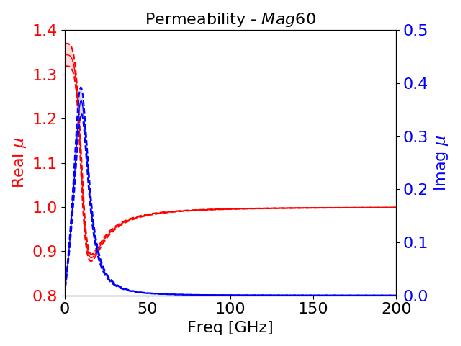}
    \caption{Dielectric permittivity and magnetic permeability for \textit{Mag60}, the real parts (\textit{Real $\epsilon$} and \textit{Real $\mu$)} are plotted in \textit{red}, while the imaginary part \textit{Imag $\epsilon$} and \textit{Imag $\mu$}) are plotted in \textit{blue}. The \textit{solid line} represents the best fit, while the \textit{shadowed area} between the \textit{dashed lines} shows the $1-\sigma$ uncertainty after the fitting routine.}
    \label{fig:mag60}
\end{figure}
\begin{figure}[htbp]
    \centering
    (a)\includegraphics[width=0.45\textwidth]{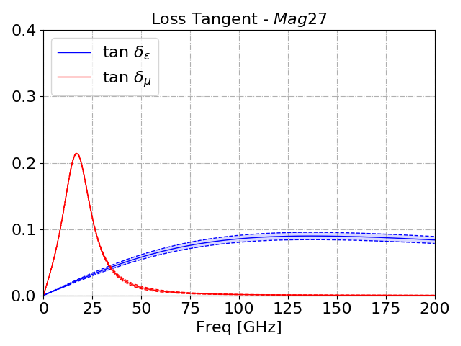}
    (b)\includegraphics[width=0.45\textwidth]{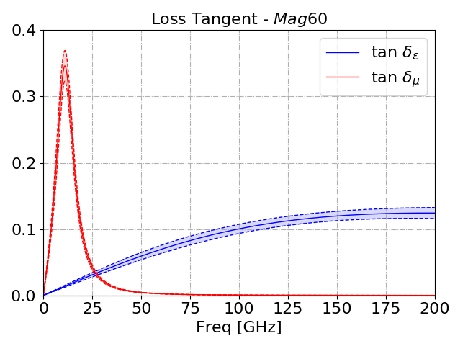}
    \caption{Dielectric (\textit{blue}) and magnetic (\textit{red}) loss tangent for \textit{Mag27} and \textit{Mag60}. The \textit{solid line} represents the best fit, while the \textit{shadowed area} between the \textit{dashed lines} shows the $1-\sigma$ uncertainty after the fitting routine.}
    \label{fig:losstang}
\end{figure}
As explained in Sec. \ref{sec:abs2}, our parametric model has five free parameters. To reduce the number of parameters to be determined through the fitting routine of the transmission data, and therefore reduce the degeneracy, we measure the static permittivity of the samples independently, using a capacitor made with two planar copper plates of area $\mathrm{A}$:
\begin{equation}\label{eq:capacitor}
        \mathrm{C=\epsilon_0\epsilon_{dc}\frac{A}{\Delta x}},
\end{equation}
where $\mathrm{C}$ is the capactitance, $\mathrm{\epsilon_0}$ is the absolute vacuum permittivity and $\mathrm{\Delta x}$ is the distance between the plates, corresponding to the thickness of the sample. By measuring the capactitance, knowing $\mathrm{A}$, $\mathrm{\Delta x}$ and $\mathrm{\epsilon_0}$, it is possible to obtain the static relative permittivity of the sample. With this preliminary step we reduce the free parameters from five to four. In Table \ref{tab:samples}, a summary of the results for the static relative permittivity and thickness of the two samples is given.

Transmission and reflection measurement are carried out using six different pairs of standard rectangular horn antennas, between $7$ and $75$ GHz, and two pairs of circular corrugated horn antennas, from $75$ up to $170$ GHz. In total we use eight pairs of antennas to cover the whole frequency range with a vector network analyzer (Agilent PNA-X N5245A with N5261A mm-wave test set and related OML extensions). We place the sample under test on the aperture of the transmitting antenna to carry out both transmission ($\mathrm{S_{21}}$) and reflection ($\mathrm{S_{11}}$) measurements at the same time.
To calibrate the data we measure $\mathrm{S_{21}}$ without the sample (free-space), and $\mathrm{S_{11}}$ for a perfectly reflective surface (mirror) and for a perfectly absorptive surface (pyramidal foam with $\mathrm{S_{11}}<-50$ $\mathrm{dB}$). The calibrated signals are:
\begin{equation}\label{eq:S21}
    \mathrm{S_{21,cal} = \frac{S_{21,sample}}{S_{21,free}}},
\end{equation}
\begin{equation}\label{eq:S11}
    \mathrm{S_{11,cal} = \frac{S_{11,sample}-S_{11,foam}}{S_{11,mirror}-S_{11,foam}}}.
\end{equation}

Using the parametric function described in Sec. \ref{sec:abs2}, we fit $\mathrm{S_{21,cal}}$ data using a least-square method in two steps:
\begin{itemize}
    \item First we exclude data below $50$ GHz, we fix $\mathrm{\epsilon_{dc}}$ using the value obtained with the capacitor measurement described previously (see Table \ref{tab:samples}), and $\mu$ to be constant and equal to unity. We use this approximation because Eq. \ref{eq:5} tends to 1 for $\nu\rightarrow\infty$. With this first step we retrieve $\mathrm{\epsilon_{\infty}}$ and $\mathrm{\nu_{\epsilon}}$. 
    \item We fit again $\mathrm{S_{21,cal}}$ data, fixing the three known parameters and including the whole frequency range from 7 to 170 GHz, to retrieve the final two parameters of our model: $\mathrm{\mu_{0}}$ and $\mathrm{\nu_{\mu}}$.
\end{itemize}

Results for the fitting routine are shown in Fig. \ref{fig:mag27_trans3} for {\it Mag27} and in Fig. \ref{fig:mag60_trans3} for {\it Mag60}. The results have been split in three panels for clarity. The upper panel shows data measured in X, Ku, K, Ka and Q bands, the central panel contains data of the measurements performed in V and W bands, and finally the lower panel shows the data of D band.  

\subsection{\label{sec:abs3}PARAMETERS VALIDATION}
After extracting all the parameters of our model using transmission data, we calculate reflection from a slab of material with the same thickness of our sample, and we compare the computed response to $\mathrm{S_{11,cal}}$ data to validate our results. Results can be seen in Fig. \ref{fig:mag27_ref3} for {\it Mag27} and in Fig. \ref{fig:mag60_ref3} for {\it Mag60}. We can see how the data in the range 18-26 GHz do not agree well with the computed one. This is likely due to an imperfect calibration in this band, however we highlight here that this mismatch doesn't have an impact on the result of the fit because we do not use reflection data to constrain the parameters of the model, but purely to cross-check the results.

Once the free parameters of the model are obtained, we can use them in Eq. \ref{eq:4} and Eq. \ref{eq:5} to compute the complex dielectric permittivity and magnetic permeability of the two materials. The results, for the materials analyzed in this work, are shown in Fig. \ref{fig:mag27} for \textit{Mag27} and in Fig. \ref{fig:mag60} for \textit{Mag60}; from these it can be seen that both values of the real and the imaginary part of both permittivity and permeability increase with the relative amount of magnetite powder in the composite material. 

As it appears clear from these data, the loss tangent is dominated by magnetic losses at low frequency. At higher frequency the magnetic loss tangent becomes negligible and the material is dominated by dielectric losses. To better illustrate this, we show the loss tangents in Fig. \ref{fig:losstang}a for \textit{Mag27} and Fig. \ref{fig:losstang}b for \textit{Mag60}. 
\newline

\section{\label{sec:res}CONCLUSIONS}
We fabricated and tested the electromagnetic properties of two samples made with Stycast 2850FT and $\mathrm{Fe_{3}O_{4}}$ powder (magnetite). The first sample has a powder-epoxy mass-ratio of $27\%$ , while the second has a mass-ratio of $60\%$. While Stycast 2850FT has already been used extensively for similar applications, most applications use CIP (carbonyl iron powder)\cite{Yang2009} as the filler. Formulas have been proposed and tested to predict the behaviour of CIP composites knowing the permittivity and permeability of the filler and the dielectric matrix and their volume ratio\cite{Merrill}. We expect similar formulas to be applicable to magnetite composites, however we measured only two samples with two different ratios, and therefore we can not validate at this stage a reliable method to predict the behaviour of Stycast-magnetite composites.

An application using Stycast 2850FT with magnetite has not been previously reported in the literature. Magnetite has been used in combination with other materials\cite{Fe3O4_1} and measurements of electromagnetic properties of these composite materials can be found in the literature\cite{Fe3O4_2, Fe3O4_3, Fe3O4_4}; however most measurements have been performed below $\sim 20$ GHz. These works showed already the possibility to tune the electromagnetic properties (specifically complex permeability and therefore the magnetic loss tangent) by changing the ratio of $\mathrm{Fe_3O_4}$ at low frequency ($\leq 20$ GHz). 

We presented here the analysis procedure and results for the measurement of dielectric permittivity and magnetic permeability of the two samples we made with magnetite powder and Stycast 2850FT. This work shows that the material can be customized for the specific application to achieve the required permittivity and permeability values by changing the $\mathrm{Fe_{3}O_{4}}$-Stycast 2850FT mass-ratio. However, as can be seen in Fig. \ref{fig:mag27}, \ref{fig:mag60} and \ref{fig:losstang}, the ability to change absorption by increasing the relative fraction of magnetite particles is limited to low frequency ($\leq 20-30$ GHz), while at higher frequencies increasing the relative ratio of magnetite does not increase neither magnetic nor dielectric losses significantly. We can notice from Fig. \ref{fig:losstang} that the dielectric loss tangent is similar between the two samples despite the different mass ratio. This is due to the fact that magnetite is a ferrimagnetic material, and as such it is characterized by low conductivity, implying that the imaginary part of the permittivity is negligible compared to the the real part. This is the reason why the dielectric loss tangent remains stable when we increase the amount of magnetite powder in the material. The ability to tune the complex permeability at low frequencies is in good agreement with results in the literature, however some applications might require knowledge of the complex permittivity and permeability on a broader frequency range for design purposes. 

The observed greater change in the imaginary part of the permeability compared to its real part can be explained because of the resonant behaviour of magnetite particles dispersed in a dielectric medium. We expect large variations of dc permeability by changing the volume ratio. However, with increasing frequency the real part of the permeability is expected to drop, while around the resonance frequency we expect mainly variation of the imaginary part of the permeability. 
A limiting factor of this analysis method is the extrapolation of the dc permeability from data at higher frequencies ($\mathrm{\nu>7}$ GHz). For future work we can imagine a set-up to measure this parameter independently similarly to what we have done with the capacitance measurements, to obtain a better constraint of this parameter and reduce even further the number of free parameters to be fitted.

In conclusion, this work shows the possibility of using Stycast 2850FT in combination with magnetite powder to fabricate an RF-absorbing material, and it describes a method to measure complex permittivity and permeability on a broad frequency range with transmission and reflection measurements. We find that increasing the relative fraction of magnetite particles impacts mainly magnetic losses at low frequency ($\leq 20-30$ GHz).

\section*{ACKNOWLEDGMENTS}
TG acknowledges the Department of Physics of the University of Oxford and Kavli IPMU for a joint fellowship to support his doctoral studies. Kavli IPMU is supported by the World Premier International Research Center Initiative (WPI), MEXT, Japan. We thank Mr. Francesco Cavaliere (University of Milano) for his help in fabricating the mould, and Dr. Robert Watkins (University of Oxford) for useful and productive discussion and comments on the manuscript. The authors are grateful to the anonymous referees for useful suggestions and discussions.
\bibliography{aipsamp}

\end{document}